\documentstyle[aps,prl]{revtex} 
\input epsf.sty 

\begin{document}
\renewcommand{\theequation}{\arabic{equation}} 

\twocolumn[ 
\hsize\textwidth\columnwidth\hsize\csname@twocolumnfalse\endcsname 
\draft 

\title{Three Phases in the 3D Abelian Higgs Model with Nonlocal 
Gauge Interactions}

\author{Shunsuke Takashima,$^1$ 
Ikuo Ichinose,$^1$  Tetsuo Matsui,$^2$ and  
Kazuhiko Sakakibara$^3$ } 
\address{$^1$Department of Applied Physics,
Nagoya Institute of Technology, Nagoya, 466-8555 Japan 
}
\address{$^2$Department of Physics, Kinki University, 
Higashi-Osaka, 577-8502 Japan 
}
\address{$^3$Department of Physics, Nara National College of Technology, 
Yamatokohriyama, 639-1080 Japan
} 

\date{\today}

\maketitle 

\begin{abstract}   
We study the phase structure of the 3D nonlocal compact
U(1) lattice gauge theory coupled with a Higgs field 
by Monte Carlo simulations. 
The nonlocal interactions among gauge variables are along 
the temporal direction and mimic the effect of local
coupling to massless particles.
In contrast to the 3D local abelian Higgs model having only 
the confinement phase, the present model exhibits
the confinement, Higgs, and Coulomb phases separated by  three
second-order transition lines emanating from a triple point.
This suggests that electron 
fractionalization phenomena in strongly-correlated electron 
systems may take place not only in the Coulomb phase but also
in the Higgs phase.

\end{abstract} 
\pacs{} 
]

\setcounter{footnote}{0} 

Some recent experiments in strongly-correlated electron systems
indicate breakdown of Landau Fermi liquid theory.
Also, for certain class of quantum phase transitions,
it is argued that Ginzburg-Landau theory is not
applicable\cite{qpt}.
In these cases, the crucial point is the change of 
particle picture:
quasiparticles are no longer electrons but certain new objects 
carrying fractional/exotic quantum numbers.
These objects may be ``constituents" of the original 
electrons. 
For example, each electron in a high-$Tc$ cuprate may be
a composite of a holon and a spinon.
Anderson\cite{anderson} argued that the anomalous properties of 
the normal state may be understood
by regarding these holons and spinons as 
quasiparticles. 
This phenomenon of dissociation
of electrons is called the charge-spin separation (CSS).
Another example is Jain's composite-fermion picture 
of the fractional quantum Hall effect (FQHE)\cite{FQHE}, where
each electron is viewed as a composite of a so-called composite fermion 
and fictitous magnetic fluxes. 
Jain argued that when these fluxes cancel the external magnetic
field partly, the FQHE is understood as the integer QHE of 
composite fermions
in the residual field.

To investigate these possibilities of fractionalization 
of electrons, knowledge and methods of gauge theory are useful. 
In fact, in path-integral formalism, one can introduce
a U(1) gauge field as an auxiliary field that 
binds the constituents of 
each electron, and  study the dissociation phenomena as a 
confinement-deconfinement phase transition (CDPT) 
of this gauge dynamics.
For the CSS, by starting from the $t-J$ model and using the hopping 
expansion, we have argued that
a CDPT takes place at certain critical temperature $T=T_{\rm CSS}$
and the CSS is possible {\it below} $T_{\rm CSS}$\cite{css}.
For the FQHE, by introducing a fermion (we called it chargon) 
for a composite fermion and a boson (fluxon) for fluxes, 
we have again supported the CDPT \cite{pfs} at $T=T_{\rm PFS}$
below which chargons and fluxons separate. We called 
this phenomenon particle-flux separation (PFS). 
Furthermore, when fluxons Bose condense, 
Jain's idea of FQHE is realized.

For the three-dimensional (3D) compact U(1) gauge system with local 
interactions {\em without coupling to matter fields}, Polyakov showed
that the system stays always in the confinement phase
due to the instanton condensation\cite{polyakov}.   
However, the issue of whether a CDPT is possible for a 3D 
(spatial 2D at $T=0$)U(1) gauge system {\it with matter fields} is still 
controversial\cite{cdpt}, mainly due to the lack of convincing
methods for investigating a system with {\it massless(gapless)} matter fields
whose effect is crucial for the phase structure.
In the previous Letter\cite{AIMS} we introduced a 3D 
compact U(1) pure lattice gauge theory with {\em nonlocal gauge 
interactions} along the temporal direction which simulate 
the effect of gapless matter fields in the parity-symmetric phase\cite{parity}. 
By numerical simulations, 
we found that the CDPT occurs at a certain gauge coupling. 
The existence
of the deconfinement phase supports
the fractionalization phenomena like the CSS and PFS.

To push forward  this gauge theoretical approach in more realistic 
manner, one needs to include another matter field, i.e.,
a Higgs field, coupled to the gauge field locally.
Actually, in the slave-boson (SB) t-J model, there appear 
fermionic spinons,
bosonic holons, and a U(1) gauge field. In the flux state
of the mean-field theory \cite{mft}, spinons are regarded as 
massless Dirac fermions in  a parity doublet. 
Such a spin system is studied also as an 
algebraic spin liquid\cite{ASL}. 
To examine the effect of these Dirac fermions on the phase structure, etc., we perform path integration 
over the fermions formally, and approximate
the resulting fermionic determinant by the nonlocal gauge 
interactions introduced in Ref.\cite{AIMS}.
To describe the bosonic holons one may introduce a U(1) Higgs field 
with the fundamental charge,
its amplitude being fixed as $\sqrt{\delta}$, where $\delta$ is
the holon density.   
In the chargon-fluxon model of the FQHE, integration over
fermionic chargons certainly generate nonlocal gauge interactions,
and bosonic fluxons may be treated as a Higgs field
coupled to this gauge field.
         
In this Letter we extend the previous study\cite{AIMS} 
by coupling a U(1) Higgs field to the gauge field locally, 
and investigate the phase structure of this
nonlocal Abelian Higgs model (AHM).
Since the ordinary (local) 
3D AHM in the London limit (no radial fluctuations) 
has only the confinement phase\cite{AHM,janke},
it is interesting to see if the nonlocal interactions among
the gauge field are capable of
generating a phase transition into the Higgs phase. 
In the SB t-J model and an algebraic spin liquid with doped holes,
the possible Higgs phase implies  a superconducting state.
In the FQHE regime, the Higgs phase is indispensable 
to achieve Jain's idea\cite{FQHE} since
a condensation of fluxons is necessary to cancel the uniform 
external magnetic field partly.

We consider the nonlocal AHM defined on the cubic lattice
of the size $V=N_0 N_1 N_2$ with the periodic boundary condition.
The compact U(1) gauge field 
$U_{x\mu}=\exp (i\theta_{x\mu}) \; (-\pi < 
\theta_{x\mu} \le \pi)$
is put on each link $(x,\mu)\equiv(x, x+\hat{\mu}) \; (\mu=0,1,2)$,
and the Higgs field $\phi_x=\exp (i\varphi_x)\;
(-\pi<\varphi_x \le \pi)$ on each site $x$.
The partition function $Z$ is given by
\begin{eqnarray}
Z&=&\int \prod_{x,\mu} dU_{x\mu}\prod_x 
d\phi_x \exp (A),\ \ \ A = A_{\rm G} + A_{\rm H}. 
\label{z}
\end{eqnarray}
$A_{\rm G}$ is the gauge part of the gauge-invariant
action $A$\cite{PL},
\begin{eqnarray}
A_{\rm G} &=& g\sum_{x}\sum_{i=1}^2
\sum_{\tau=1}^{N_0} c_{\tau}(V_{x,i,\tau}
+\bar{V}_{x,i,\tau})
+A_{\rm S},\nonumber\\
V_{x,i,\tau} &=& \bar{U}_{x+\tau\hat{0},i}
\prod_{k=0}^{\tau-1}\left[\bar{U}_{x+k\hat{0},0}
U_{x+\hat{i}+k\hat{0},0}\right]U_{xi},\nonumber\\
A_{\rm S} &=& g \lambda\sum_x(\bar{U}_{x2}
\bar{U}_{x+\hat{2},1}U_{x+\hat{1},2}U_{x1}+ {\rm c.c.}),
\end{eqnarray}
where $g$ is the (inverse) gauge coupling constant,  
$V_{x,i,\tau} $ 
is the product of $U_{x\mu}$ along the rectangular
$(x, x+\hat{i}, x+\hat{i}+\tau \hat{0},x+ \tau \hat{0})$ 
of the size $(1 \times \tau)$ in the $(i-0)$ plane ($i=1,2$), and
$c_{\tau}$ is the  {\em ``nonlocal coupling constant"}.
$A_{\rm S}$ is the spatial single-plaquette coupling of $U_{x\mu}$'s.
$A_{\rm H}$ is the {\em local} Higgs action,
\begin{eqnarray}
A_{\rm H}&=&\kappa \sum_{x, \mu}\left(\bar{\phi}_{x+\hat{\mu}}
U_{x\mu}\phi_x  + {\rm c.c.}\right),
\end{eqnarray}
where $\kappa$ is the Higgs coupling. In the SB
$t-J$ model, $\kappa \propto \delta$ as explained.

In the previous study\cite{AIMS} of the pure gauge 
system ($\kappa=0$), we considered 
(i) power-law decay  $c_{\tau}=1/\tau$ simulating
the effect of massless relativistic matter fields and
(ii) exponential decay  $c_{\tau}=\exp(-m\tau)$
mimicking massive matter fields. 
We found a second-order CDPT for the  case (i); 
the critical coupling is 
$g_c=0.10 \sim 0.12$ for $\lambda=1$, while no CDPT for the 
case (ii). We also studied (iii) no decay case 
$c_{\tau}=$const for nonrelativistic fermions in which 
the CDPT was observed at very small coupling like $g_c=0.03$ 
for $V=24^3$. Then one can naturally expect that if the Higgs 
transition exists in the power-law decay case, it also exists in the 
no-decay case.

\begin{figure}[thbp]
\begin{center}
\leavevmode
\epsfxsize=7cm
\epsffile{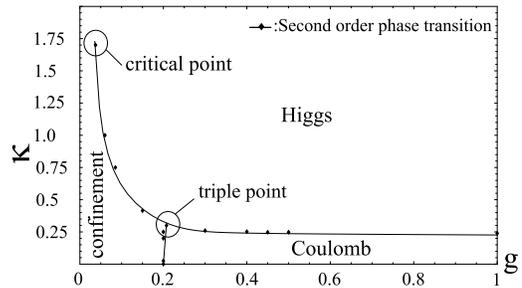}
\vspace{0.2cm}
\caption{Phase diagram of the 3D nonlocal AHM.  
There are three transition lines, which are determined 
by the location of the peak of $C$ for $N=16$. 
}
\label{fig1}
\end{center}
\end{figure}


In the following numerical studies of the gauge model Eq.(\ref{z}), 
we concentrate on the 
power-law decay case (i) 
with  $\lambda=1$ on the cubic lattice $V=N^3$ up to $N=24$.
We first measure the ``internal energy" $E$ and
the ``specific heat" $C$ defined as
\begin{eqnarray}
E &\equiv& -\frac{1}{V}\langle A \rangle,\ \ \
C \equiv \frac{1}{V}\langle (A- \langle A \rangle)^2 \rangle.
\label{EC}
\end{eqnarray}

In Fig.\ref{fig1} we present the phase diagram in the 
$g-\kappa$
plane determined by the measurement of $E$ and $C$. 
There are {\em three transition lines} separating 
three phases which we call 
the confinement, Higgs, and Coulomb phases.
(We shall identify each phase later 
by measuring the instanton density and the Higgs boson mass.)

Let us first consider the confinement-Coulomb transition line.
It includes the CDPT at $\kappa=0$ previously
found in Ref.\cite{AIMS}.
Thus the deconfinement phase observed in Ref.\cite{AIMS} is
identified as the Coulomb phase.
We measured $E$ and $C$ along a fixed $\kappa$ and
found that $E$ exhibits no hysteresis and 
the peak of $C$ develops as the system size $N^3$ increases 
just as in the $\kappa=0$ case\cite{AIMS}. 
These are the typical signals of a second-order phase transition.   
We note that the location of the peak of $C$ shifts 
into the smaller $g$  (higher-$T$) direction for larger $N$, 
just the opposite to the case of local interaction.
It is because the new gauge interactions are
introduced as $N$ increases, and these terms favor to suppress fluctuations of 
$U_{x\mu}$\cite{AIMS}.
 
\begin{figure}[htbp]
\begin{center}
\leavevmode
\epsfxsize=7cm
\epsffile{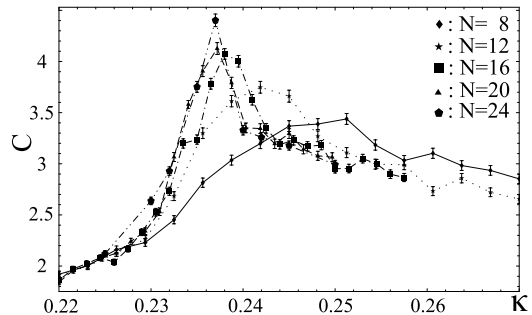}
\vspace{0.3cm}
\caption{Specific heat $C$ vs. $\kappa$ 
for $g=1.0$. The peak develpes as $N$ increases.
}
\label{fig2}
\end{center}
\end{figure}

Next, let us consider the Coulomb-Higgs transition.
In Fig.\ref{fig2} we present $C$ vs. $\kappa$ for $g=1.0$.
As $N$ increases,  

\begin{figure}[htbp]
\begin{center}
\leavevmode
\epsfxsize=6.5cm
\epsffile{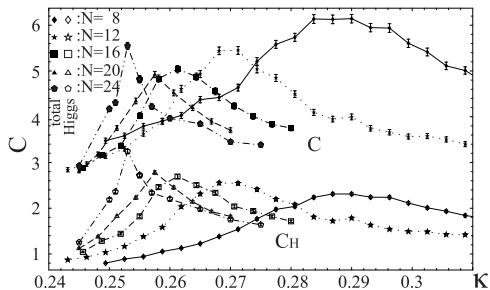}
\vspace{0.3cm}
\caption{Total specific heat $C$ (upper) and
its Higgs part $C_{\rm H}$ (lower) vs. $\kappa$ for $g=0.3$.}
\label{fig3}
\end{center}
\end{figure}

\noindent
the  peak of $C$ also develops. 
Again, the location of the peak shifts to smaller
$\kappa$ because of the nonlocal action $A_{\rm G}$.   
To confirm that this transition is of second order,
we apply the  finite-size scaling hypothesis\cite{FSS}
to $C$ of Fig.\ref{fig2}. 
We fit $C$ as a function of  $N$ 
and $\epsilon\equiv (\kappa-\kappa_\infty)/\kappa_\infty$, 
where $\kappa_\infty$ is the critical Higgs coupling 
at $N\rightarrow \infty$, in the form 
of $C_N(\epsilon)=N^{\sigma/\nu}
\phi(N^{1/\nu}\epsilon)$ with certain 
scaling function $\phi(x)$.
We get a quite good fitting for $N=8 \sim 24$
with $\nu=0.75, \kappa_\infty=0.23$ and $\sigma=0.12$.


In Fig.\ref{fig3} we present $C$ for a smaller value
of $g$, $g=0.3$.
As $N$ increases from $N=8$ to $20$, $C$ gets sharper but 
the height of its peak becomes lower. 
To understand this interesting behavior, we measured the 
``specific heats" of the gauge part and Higgs part,
$C_{\cal O}=\langle (A_{\cal O}- \langle A_{\cal O}\rangle)^2
\rangle/V$ (${\cal O}=$G and H),  
 separately\cite{janke}.
$C_{\rm H}$ in Fig.\ref{fig3}
has a developing peak as in the usual second-order 
phase transition.
On the other hand, both $C_{\rm G}$ and the cross term,
$C-C_{\rm H}-C_{\rm G},$
are smooth and reduce as $N$ increases.
As explained above, the gauge specific heat $C_{\rm G}$ 
has the peak which shifts to left (smaller $g$) 
for increasing $N$. 
Then the tail of $C_{\rm G}$ observed at a fixed 
$g$  on the right of the confinement-Coulomb transition line
 reduces as $N$ increases. 
This reduction gives rise to the above behavior of $C$
 in Fig.\ref{fig3} at $g=0.3$ which is 
close enough to the transition line  
 at $g \sim 0.20$ (See Fig.\ref{fig1}).
For larger $N ( \ge 24)$, the peak of $C$
starts to develop as it should be since the Higgs part,
the relevant part of the Coulomb-Higgs transition, starts
to dominate over $C_{\rm G}$ and the cross term.

\begin{figure}[htbp]
\begin{center}
\leavevmode
\epsfxsize=7cm
\epsffile{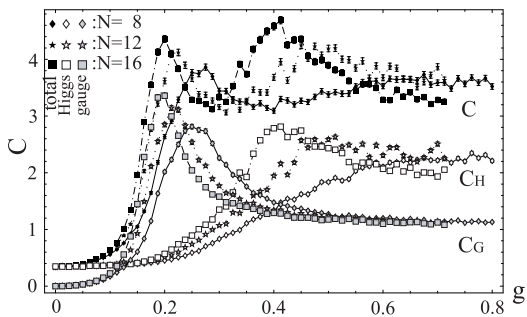}
\vspace{0.3cm}
\caption{Specific heat $C, C_{\rm H}, C_{\rm G}$
vs. $g$  for $\kappa=0.25$. $C$ has two peaks
reflecting each peak of $C_{\rm H}$ and $C_{\rm G}$.}
\label{fig4}
\end{center}
\end{figure}

\begin{figure}[htbp]
\begin{center}
\leavevmode
\epsfxsize=7cm
\epsffile{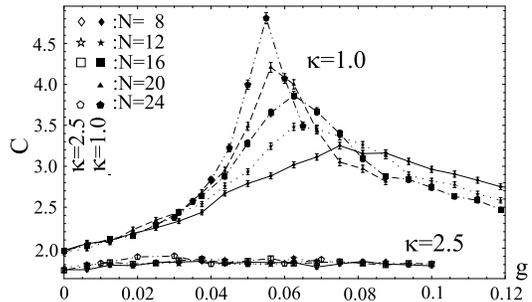}
\vspace{0.3cm}
\caption{$C$ vs. $g$. For $\kappa=1.0$,
$C$ has a peak that develops as $N$ increases.
For $\kappa=2.5$, $C$ has no peak.
}
\label{fig5}
\end{center}
\end{figure}

Near the triple point, both $C_{\rm G}$ and $C_{\rm H}$ are 
relevant. 
In Fig.\ref{fig4} we present $C$ vs. $g$
for $\kappa=0.25$. We choose $\kappa=
0.25$ as a horizontal line because it crosses the two
 transition lines near the triple point. 
Each peak of $C_{\rm G}$ and $C_{\rm H}$
develops with $N$ simultaneously as expected. 

Next, let us consider the confinement-Higgs transition.
In Fig.\ref{fig5} we present $C$ vs. $g$ for
$\kappa=1.0$.
At $g=0$, one can exactly integrate over $U_{x\mu}$ to obtain
$Z=[I_0(2\kappa)]^{3V}$, where $I_0$ is the modified 
Bessel function. Thus $Z$ is an analytic function of $\kappa$,
showing that there are no  transitions along the line $g=0$.
We measured $E$ and $C$ at several fixed $\kappa$'s 
by varying $g$ to find that the signal of 
\noindent
second-order 
phase transition is getting weaker as $\kappa$ increases.
The phase transition line seems to terminate  
at the critical point, the location  of which is roughly 
estimated as $g=0.03, \kappa=1.75$.
This critical point is viewed as a nonlocal version of the 
complementarity between the Higgs and confinement phases
discussed in Ref.\cite{complementarity}.

\begin{figure}[htbp]
\begin{center}
\leavevmode
\epsfxsize=4.2cm
\epsffile{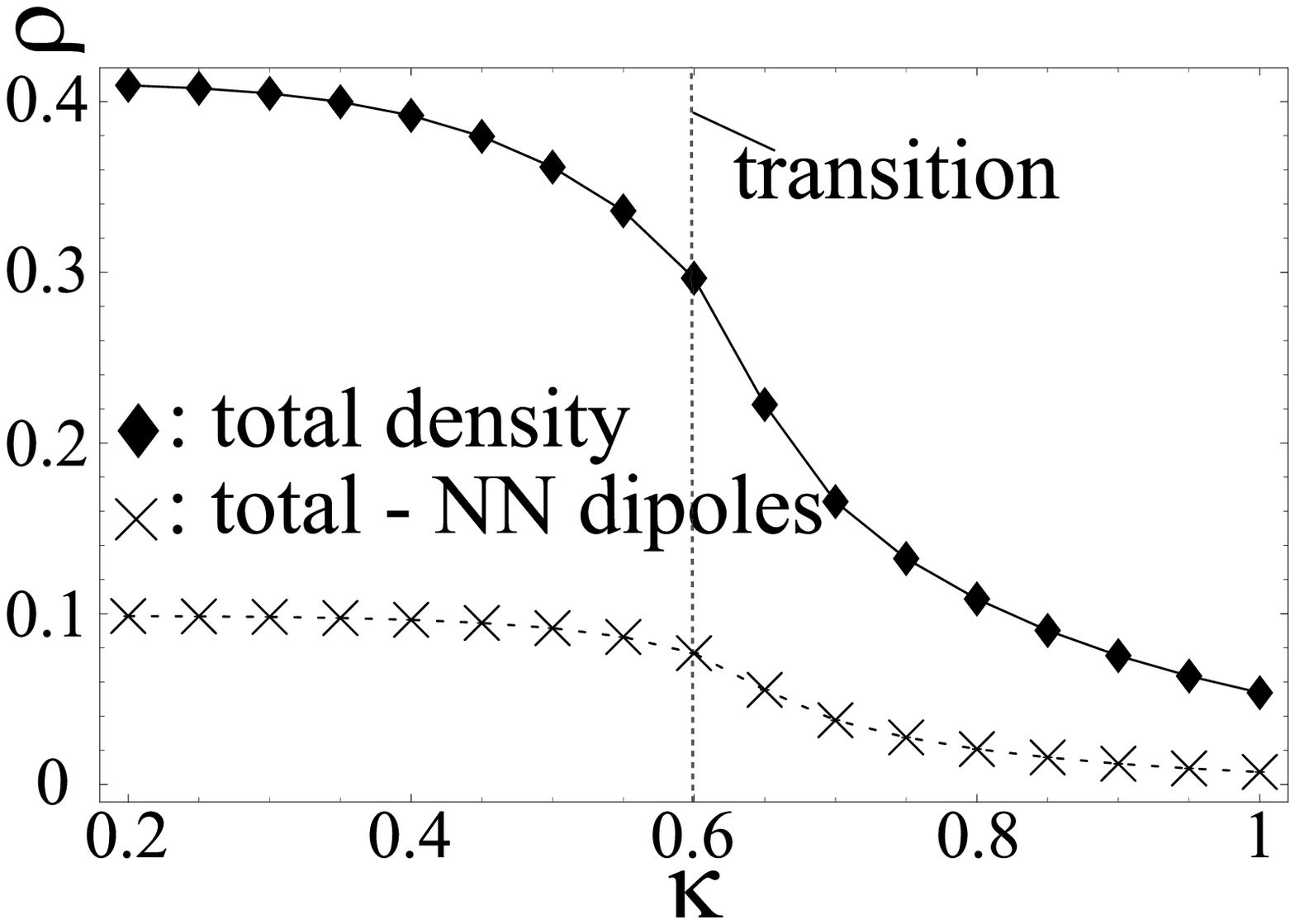}
\epsfxsize=4.2cm
\epsffile{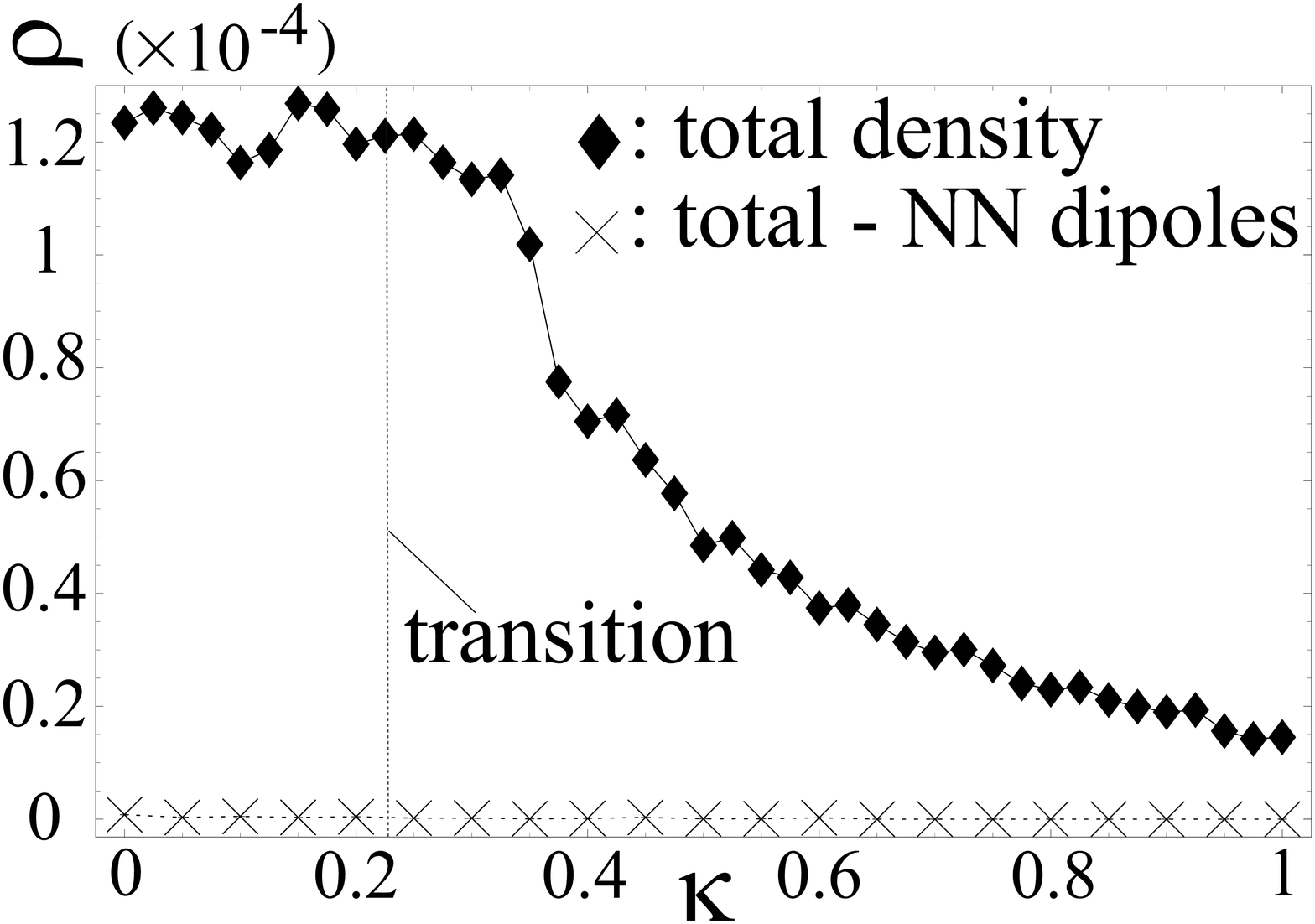}
(a) $g=0.1$ \hspace{3cm} (b) $g=0.5$
\vspace{0.3cm}
\caption{Instanton density $\rho$ for $N=16$ vs. 
$\kappa$ for (a) $g=0.1$ 
and (b) $g=0.5$. The crosses are the density
$\rho_{\rm single }$ of isolated instantons. 
Dashed lines indicate the transition points. 
$\rho(g=0.5)$ is smaller than $\rho(g=0.1)$ by
the factor $\sim 10^{-4}$.
}
\label{fig6}
\end{center}
\end{figure}

In order to characterize these 
three phases, let us calculate the instanton density 
$\rho_x$\cite{instanton}.
$\rho_x$ is the strength of monopoles sitting on the site of the 
dual lattice $x+(\hat{1}+\hat{2}+\hat{3})/2$, which 
measures the disorderedness of gauge-field configurations.  
In Fig.\ref{fig6} we show the average density 
$\rho \equiv \langle |\rho_x| \rangle $ for $g=0.1$ and $g=0.5$.
In Fig.\ref{fig6}(a), $\rho$ is fairly large  
at the small $\kappa$ region, and as crossing the transition point, 
it decreases rapidly.
In Fig.\ref{fig6}(b), $\rho$ is already very low 
even in the small $\kappa$'s, and it 
decreases further across the transition point.
Such behavior of $\rho$ and the general 
relations $\rho\,({\rm confinement})> \rho\,({\rm Coulomb})
 > \rho\,({\rm Higgs})$ lead us to identification of 
each phase as shown in Fig.\ref{fig1}.

For the AHM, one may study vortices of 
the Higgs field and their 
trajectories which form closed loops
or terminate at (anti-)instantons\cite{vortex}.   
The confinement phase  is 
a plasma phase of instantons\cite{polyakov}, i.e.,
the CDPT is characterized by dissociations
of instanton-anti-instanton dipoles connected
by vortex lines into a system of instanton plasma\cite{vortex}.
In Fig.\ref{fig6} we also plot the density  of
isolated instantons $\rho_{\rm single}$ 
by subtracting the nearest-neighbor 
dipole instanton  pairs in the temporal direction.
Fig.\ref{fig6} clearly shows that $\rho_{\rm single}$ 
survives only in the confinement phase as it should be.

\noindent

\begin{figure}[htbp]
\begin{center}
\leavevmode
\hspace{-1.5cm}
\epsfxsize=9cm
\epsffile{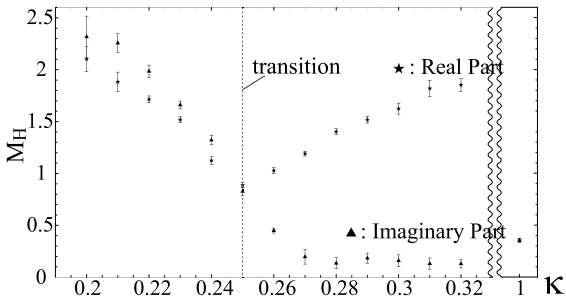}
\vspace{0.3cm}
\caption{Masses $M_{\rm H}(N=16)$ of real and imaginary parts of 
$(\bar{\phi}_{x+\mu}U_{x\mu}\phi_x)$ vs. $\kappa$ for $g=0.4$. 
$M_{\rm H}$ of the real part has its minimum at
the transition point as in the 4D local AHM.
}
\label{fig7}
\end{center}
\end{figure}

Finally, let us consider masses 
of gauge-invariant operators.
In Fig.\ref{fig7}, we show the masses 
$M_{\rm H}$ (inverse correlation length) of 
Re$(\bar{\phi}_{x+\mu}U_{x\mu}\phi_x)$ and 
Im$(\bar{\phi}_{x+\mu}U_{x\mu}\phi_x)$ vs. $\kappa$ for 
$g=0.4$\cite{TIM}.
The two $M_{\rm H}$ of Fig.\ref{fig7} exhibit
the behavior across the transition point
similar to that of the 4D local AHM\cite{QED4}. 
We note that the 4D local AHM has a phase structure
similar to the present model;
three phases, a triple point and the complementarity.
However at present, the  confinement-Higgs 
transitions in the 4D AHM is believed to be of
first order, which is in contrast to the
present model.

In conclusion, we numerically studied the 3D nonlocal AHM and found
that the nonlocal interactions generate all the known
phases of gauge theory. 
As explained, the existence of 
the Higgs phase supports both the superconducting state
of the $t-J$ model and the FQHE
as electron  fractionalization phenomena like CSS and 
PFS.
In fact, the  high-$T_c$ cuprates near $T=0$ enter into
the superconducting state {\it above} some critical
doping $\delta_c$. This is interpreted in Fig.\ref{fig1}
 as entering into the Higgs phase as $\kappa$ increases since
$\kappa \propto \delta$.
For a parity-preserving QED$_3$ 
coupled with a Higgs field, 
the result of the present model implies that a Higgs phase
transition occurs for a sufficiently large number of fermion flavors
since $g$ is supposed to be proportional to the number
of massless relativistic fermions.

\vspace{-0.5cm}



\begin{references}
\bibitem{parity}
For Dirac fermions without parity invariance, 
the Chern-Simons term may be generated.
See Sect.2.3 of the  second reference of Ref.\cite{AIMS}.
\bibitem{PL} 
Strictly speaking, a term of  Polyakov 
lines, $A_{\rm PL}=g
\sum_{x,i,\tau}c_{N_0}(P_x+ \bar{P}_x),\ 
P_x\equiv \prod_{\tau}U_{x+\tau\hat{0},0}$ is to be added to $A_{\rm G}$.
This term works as an external field to violate the global
symmetry $\theta_{x0}\rightarrow \theta_{x0}+\alpha$.
We performed separate simulations and checked that inclusion of this term does 
not affect the phase structure presented in the present paper.
\end{references}
\end{document}